\documentclass[12pt]{article}
\usepackage{epsfig}
\title{Quadratic integrals of motion for the systems of identical particles}
\author{Y. Brihaye\\
Faculte des Sciences,\\ Universite de Mons, 7000 Mons, Belgium\\
C. Gonera\thanks{supported by KBN grant 5 P03B06021},  
P. Kosi\'nski$^*$, P. Ma\'slanka$^*$\ 
\\Department of Theoretical Physics II \\
University of {\L}\'od\'z \\
Pomorska 149/153, 90 - 236 {\L}\'od\'z/Poland\\
S. Giller\thanks{supported by KBN grant, No. 5PO3B 06021}\\
Pedagogical University of Czestochowa,\\
 Armii Krajowej 13/15, 42-200 Czestochowa Poland.}
\date{}

\begin{document}
\maketitle
\begin{abstract}
The dynamical systems of identical particles admitting quadratic integrals of motion are classified.
The relevant integrals are explicitly constructed and their relation to separation of variables
in H-J equation is clarified.
\end{abstract}

\newpage
\section{Introduction}

Few years ago Braden \cite{b1} classified all translationally invariant systems of identical particles admitting an additional
(apart from energy and total momentum) integral of motion, which is polynomial in momenta of degree three (actually, an additional
assumption in Braden's paper is that the third-order term is q - independent). It appeared that the assumption of $S_N$\ symmetry 
put the severe restrictions on the admissible hamiltonians.\\
In the present paper we classify the dynamical systems of identical particles possesing the quadratic integrals of motion.
We find the general form of potentials and we show that the additional integrals arise from the separation
of radial variable in Jacobi coordinates. We explain also the group - theoretic origin of the
integrals of motion.
\section{Quadratic integrals of motion}

We want to classify all classical dynamical systems of $N$\ degrees of freedom obeying the following assumptions:\\
(i) the hamiltonian takes the natural form
\begin{eqnarray}
H=\frac{1}{2}\sum_{i=1}^{N}p_i^2+V(q_1,\;\ldots ,\;q_N) \label{w1}
\end{eqnarray}
(ii) $H$\ is translationally invariant
\begin{eqnarray}
\{H,\;P\}=0,\;\;P=\sum_{i=1}^Np_i\label{w2}
\end{eqnarray}
(iii) $H$\ is invariant under the action of the group $S_N$\ of all permutations of canonical
variables $q_i,\;p_i,\;i=1,\ldots,\;N$.\\
(iv) $H$\ admits at least one integral of motion quadratic in momenta and functionally independent of
$H$\ and $P$. \\
To this end we make the following observations:\\
a) the hamiltonian (\ref{w1}) is time - reversal invariant: $q_i\rightarrow\;q_i,\;p_i\rightarrow \;-p_i,\;i=1,
\;\ldots,\;N$, is the symmetry of $H$. Therefore, one can assume from the very begining that the 
integrals of motion are either even or odd functions of momenta. In particular, our quadratic integral can
be written as
\begin{eqnarray}
I(\underline{q},\;\underline{p})=\tilde{I}(\underline{q},\;\underline{p})+W(\underline{q}),\label{w3}
\end{eqnarray}
where $\tilde{I}(\underline{q},\;\underline{p})$\ is quadratic homogeneous function of momenta.\\
b) $\tilde{I}(\underline{q},\;\underline{p})$\ is itself an integral of motion in the free case $(V\equiv 0)$. 
This follows by considering the highest - order term (in momenta) of $\dot{I}$. Now, for a free 
motion any integral is a function of momenta and the combinations (angular momentum components) 
$q_ip_j-q_j p_i,\;i,\;j=1,\ldots,\;N$. Consequently, $I$\ is also at most quadratic in $q$'s.\\
c) $S_N$- symmetric quadratic function $\tilde{I}(\underline{q},\;\underline{p})$\ can be written as a function
of the variables $\sum_{i=1}^Nq_i^{\alpha}p_i^{\beta},\;\alpha,\;\beta =0,\;1,\;2$.\\
Using (a) - (c) one easily finds that the integral $I$\ can be written in the following general form:
\begin{eqnarray}
&&I(\underline{q},\;\underline{p})=\alpha \left((\sum_{i=1}^Nq_i^2)(\sum_{i=1}^Np_i^2)-(\sum_{i=1}^Nq_ip_i
)^2\right)+\nonumber \\
&&+\beta \left(2QP(\sum_{i=1}^Nq_ip_i)-Q^2(\sum_{i=1}^Np_i^2)-P^2(\sum_{i=1}^Nq_i^2)\right)+\label{w4} \\
&&+\gamma \left(P(\sum_{i=1}^Nq_ip_i)-Q(\sum_{i=1}^Np_i^2)\right)+W(\underline{q})\nonumber
\end{eqnarray}
where $Q=\sum_{i=1}^Nq_i$.\\
Imposing the condition $\dot{I}=0$\ one arrives at the set of equations determining $W$. The integrability conditions for 
this set read:
\begin{eqnarray}
\left[2\alpha M_{kj}+(2\beta Q+\gamma )(\frac{\partial}{\partial q_k}-\frac{\partial}{\partial q_j})\right](D+2)V=0 ;\label{w5}
\end{eqnarray}
here
\begin{eqnarray}
&&M_{kj}=q_k\frac{\partial}{\partial q_j}-q_j\frac{\partial}{\partial q_k}\nonumber \\
&&D=\sum_{i=1}^Nq_i\frac{\partial}{\partial q_i} \label{w6}
\end{eqnarray}
are the generators of rotations and dilatations, respectively ( $j,\;k=1,\;\ldots, N$ )\\
\\
It is not difficult to find the general solution to eqs. (\ref{w5}). Keeping in mind that $V$\ is defined up to an additive constant one
finds:\\
1) if $\alpha \neq \beta N$\ or $\gamma \neq 0$\ V is a translationally invariant homogeneous function of degree $-2$.\\
2) if $\alpha = \beta N$\ and $\gamma =0$\ then 
\begin{eqnarray}
V(\underline{q})=\tilde{V}(\underline{q})+U\left(\sum\limits_{i=1 \atop j=1}^N(q_i-q_j)^2\right),\label{w7}
\end{eqnarray}
where $\tilde{V}$\ is a translationally invariant homogeneous function of degree $-2$\ while $U$\ is an arbitrary differentiable
function of one variable.\\
Having determined $V$ one can find $W$. The result are as follows.\\
In the case (1) any quadratic integral is a linear combination of the following ones:
\begin{eqnarray}
&&I_1=\left(\sum_{k=1}^Nq_kp_k\right)P-2QH\nonumber \\
&&I_2=2\left(\sum_{k=1}^Nq_k^2\right)H-\left(\sum_{k=1}^Nq_kp_k\right)^2 \label{w8} \\
&&I_3=2QP\left(\sum_{k=1}^Nq_kp_k\right)-2Q^2H-\left(\sum_{k=1}^Nq_k^2\right)P^2\nonumber
\end{eqnarray}
Note that $V$\ enters the above expressions only through $H$. The integrals $I_1,\;I_2,\;I_3$\
are, however, not functionally independent. Instead, they obey the relation
\begin{eqnarray}
I_1^2+P^2I_2+2HI_3=0\label{w9}
\end{eqnarray}
In the case (2), when $V$\ is given by eq. (\ref{w7}) with nontrivial $U$, there exists one independent quadratic integral:
\begin{eqnarray}
&&I_4=2\left(N(\sum_{k=1}^Nq_k^2)-Q^2\right)\tilde{H}-N(\sum_{k=1}^Nq_kp_k)^2+\nonumber \\
&&+2QP(\sum_{k=1}^Nq_kp_k)-P^2(\sum_{k=1}^Nq_k^2)\label{w10}
\end{eqnarray}
where $\tilde{H}$\ is obtained from $H$\ through the repacement $V \rightarrow  \tilde {V} $.
Note that, in the case $U=0$, $I_4$\ can be written as $I_4=NI_2+I_3$. Therefore in this case one can take $I_1$\ and $I_4$\
as independent quadratic integrals.

\section{Separation of variables} 

Typically quadratic integrals of motion are related to the separation of variables by point transformation. Let us introduce the
Jacobi coordinates \cite {b2}
\begin{eqnarray}
&&y_0=\frac{1}{N}Q\label{w11} \\
&&y_k=\frac{1}{\sqrt{k(k+1}}(kq_{k+1}-\sum_{i=1}^kq_i),\;\;\;\;k=1,\;\ldots ,\;N-1 \nonumber
\end{eqnarray}
and, as a next step the polar coordinates $\rho,\;\theta_1,\;\ldots,\;\theta_{N-2}$\ in the space of Jacobi vatiables $y_1,\;\ldots,
\;y_{N-1}$. Then
\begin{eqnarray}
\rho^2\equiv \sum_{k=1}^{N-1}y_k^2=\frac{1}{2N}\sum_{i,\;j=1}^N(q_i-q_j)^2\label{w12}
\end{eqnarray}
Moreover, the Hamiltonian for the general potential (\ref{w7}) can be rewritten in the form
\begin{eqnarray}
H=\frac{1}{2N}P^2+\frac{1}{2}\left(p_{\rho}^2+\frac{F(\underline{\theta},\underline{p}_{\theta})}{\rho^2}\right)
+\frac{G(\underline{\theta})}{\rho^2}+U(\rho^2)\label{w13}
\end{eqnarray}
where $F$\ comes from the kinetic part while $\tilde{V}=\rho^{-2}G$. Inserting (\ref{w13}) into the Hamiltonian - Jacobi equation we find
that $H$\ and $P$\ are related (as usual) to the separation of time and center - of - mass motion while the separation of radial
variable gives the additional integral
\begin{eqnarray}
C=\frac{1}{2}F(\underline{\theta},\underline{p}_{\theta})+G(\underline{\theta})\label{w14}
\end{eqnarray}
Note that $C$\ is dilatation invariant. In fact, one readily checks that
\begin{eqnarray}
C=\frac{1}{2N}I_4\label{w15}
\end{eqnarray}
which explains the origin of our quadratic integral.\\
If $U=0$\ we have two independent quadratic integrals. In order to find the second one let us write out the solution to the H-J equation:
\begin{eqnarray}
S=\frac{1}{N}PQ+S_1(\rho)+S_2(\underline{\theta})-Et\label{w16}
\end{eqnarray}
where $S_{1,\;2}$\ obey:
\begin{eqnarray}
&&\frac{1}{2}F(\underline{\theta},\;\frac {\partial S_2}{\partial {\underline {\theta }}}) +G(\theta)=\frac{1}{2N}I_4\label{w17}\\
&&\frac{1}{2}\left(\frac{\partial S_1}{\partial \rho}\right)^2+\frac{1}{2N\rho^2}I_4+\frac{1}{2N}P^2=E\nonumber
\end{eqnarray}
The equation for $S_1$\ can be solved explicitly
\begin{eqnarray}
S_1=\pm \int d\rho \sqrt{2E-\frac{P^2}{N}-\frac{1}{N\rho^2}I_4}\label{w18}
\end{eqnarray}
Eqs. (\ref{w13}) - (\ref{w18}) imply
\begin{eqnarray}
\frac{\partial S}{\partial P}=\frac{I_1}{P^2-2NH}\label{w19}
\end{eqnarray}
However, by standard algorithm \cite{b3} $\frac{\partial S}{\partial P}$\ is a constant. We conclude that $I_1$\ arises also from
separation procedure for H-J equation.

\section{$sL({\bf 2},\;{\bf R})$\ symmetry}

The result obtained in the case $U\equiv 0$\ can be also understood from the point of view of dynamical $sL({\bf 2},\;{\bf R})$\
symmetry. Due to the fact that $V$\ is homogeneous function of degree $-2$\ the three functions $H$, $X=\sum_{k=1}^Nq_k^2$\ and $Y=\sum_{k=1}^Nq_kp_k$\
form the $sL({\bf 2},\;{\bf R})$\ algebra \cite{b4}
\begin{eqnarray}
&&\{Y,\;H\}=2H\nonumber\\
&&\{Y,\;X\}=-2X \label{w20}\\
&&\{H,\;X\}=-2Y\nonumber
\end{eqnarray}
One can define the action of this algebra in the space of functions on phase space as follows \cite{b5}:
\begin{eqnarray}
(\hat{\Lambda }f)(\underline{q},\;\underline{p})=\{\Lambda (\underline{q},\;\underline{p}),f(\underline{q},\;\underline{p})\}\label{w21}
\end{eqnarray}
where $\Lambda = X,\;Y,\;H$.
Eq. (\ref{w21}) defines the representation of the algebra (\ref{w20}). In particular, the integrals of motion provide the highest - weight
vectors. If, in addition, such an integral is a homogeneous function of natural degree (with $\underline{p}\;(\underline{q})$\
having weight $+1(-1)$) and a polynomial in $\underline{p}$, it defines a finite-dimensional representation. From this point of view 
$I_1$\ and $I_2$\ are highest - weight vectors of doublet and singlet, respectively. 
It is particularly simple to trace the origin of $I_2$; it can be written in terms of $sL({\bf 2},\;{\bf R})$\ generators as
\begin{eqnarray}
I_2=2XH-Y^2\label{w22}
\end{eqnarray}
and one  immediately recognizes that $I_2$\ is simply the Casimir function for this algebra.\\
To trace the origin of $I_1$\ let us note the following.
If $I$\ is the highest - weight vector
$(\dot{I}=\{I,\;H\}=0,\;\{Y,\;I\}=nI)$\ then the next - to - highest vector $J=\{X,\;I\}$\ obeys
\begin{eqnarray}
\dot{J}=\{J,\;H\}=\left\{\{X,\;I\},\;H\right\}=-2nI;\label{w23}
\end{eqnarray}
therefore $J$\ evolves linearly in time. \\
Now, if $I$' is a second integral of degree $n$' it follows from eq. (\ref{w23}) that $n'I'J-nIJ'$\ is also an integral of motion
\cite{b6}. Applying this reasoning to the triple $(H,\;Y,\;X)$\ and the doublet $(P,\;Q)$\ one obtains $I_1$.


\begin{thebibliography}{99}

\bibitem{b1}
 H. W. Braden, "Rigidity, Functional Equations and the Calogero-Moser Model" solv-int. /00 05046.
\bibitem{b2}
 see, for example, F. Calogero, J. Math. Phys. {\bf 12}, (1971), 419
\bibitem{b3}
 L. D.Landau, E. M. Lifshitz, Mechanics, 3-d ed. Pergamon Press, 1976.
\bibitem{b4}
 P. I. Gambardella, J. Math. Phys. {\bf 16} (1975),m 1172\\
G. Barucchi, T. Regge, J. Math. Phys. {\bf 18} (1977), 1149 \\
S. Wojciechowski, Phys Lett. {\bf A64} (1977), 273
\bibitem{b5}
 C. Gonera, P. Kosinski, Acta Phys. Pol. {\bf B30} (1999), 907
\bibitem{b6}
 S. Wojciechowski, Phys. Lett. {\bf A95} (1983), 279     
 \end{thebibliography}
\end{document}